\documentclass[fleqn,12pt]{article}
\usepackage{latexsym}
\usepackage{epsfig}
\usepackage{graphicx,psfrag}
\usepackage{amssymb}
\usepackage{times}
\usepackage{epsf}
\begin{document}

\centerline{\Large\bf Efficient Calculation of P-value and Power for Quadratic }
\centerline{\Large\bf Form Statistics in Multilocus Association Testing}
\vskip 0.5cm
\centerline{Liping Tong$^{1,2}$,  Jie Yang$^3$,  Richard S. Cooper$^2$}

\vskip 0.5cm
\centerline{\today}

\vskip 0.5cm
\noindent 1. Department of Mathematics and Statistics, Loyola University Chicago, IL 60626

\noindent 2. Department of Preventive Medicine and Epidemiology, Loyola University Medical School, Maywood, IL 60153

\noindent 3. Department of Mathematics, Statistics, and Computer Science, University of Illinois at Chicago, Chicago, IL 60607

\vskip 0.5cm
\noindent Correspondence: Liping Tong, ltong@luc.edu

\vskip 0.5cm
\noindent Emails: \\
Liping Tong, ltong@luc.edu;\\
Jie Yang, jyang06@math.uic.edu; \\
Richard S. Cooper, rcooper@lumc.edu

\newpage
\noindent {\bf\Large Abstract}

We address the asymptotic and approximate distributions
 of a large class of test statistics with quadratic forms used in association studies.
 The statistics of interest do not necessarily follow a chi-square
 distribution and take the general form $D=X^T A X$, where $X$ follows
 the multivariate normal distribution,
 and $A$ is a general similarity matrix which may or may not be positive semi-definite.
 We show that $D$ can be written as a linear combination
 of independent chi-square random variables, whose distribution can
 be approximated by a chi-square
 or the difference of two chi-square distributions.
 In the setting of association testing, our methods are especially useful in two
 situations. First, for a
 genome screen, the required significance level is much smaller
 than 0.05 due to multiple comparisons, and estimation of p-values
 using permutation procedures is particularly challenging.
 An efficient and accurate estimation procedure would therefore be useful.
 Second, in a candidate gene study based on haplotypes when phase
 is unknown a computationally expensive method-the
 EM algorithm-is usually required to infer
 haplotype frequencies.
 Because the EM algorithm is needed for each permutation,
 this results in a substantial computational burden,
 which can be eliminated with our mathematical solution.
 We assess the practical utility of our method using extensive simulation studies
 based on two example statistics and apply it to find the sample
 size needed for a typical candidate gene association study when phase information is
 not available.
 Our method can be applied to any quadratic form statistic and therefore should
 be of general interest.


\newpage
\noindent {\bf\Large Introduction}

The multilocus association test is an important tool for use in the genetic
 dissection of complex disease. Emerging evidence demonstrates that multiple
 mutations within a single gene often interact to create a ``super allele"
 which is the basis of the association between the trait and the genetic locus
 [Schaid et al. 2002].
 For the case-control design, a variety of test statistics have been applied,
 such as the likelihood ratio, $\chi^2$ goodness-of-fit, the score test,
 the similarity- or distance-based test, etc.
 Many of these statistics have the quadratic form $\hat s^T A\hat s$,
 or are functions of quadratic forms, where $\hat s$ is a function of
 the sample proportions of haplotype or genotype categories and $A$
 is the similarity or distance matrix.  Some of these test statistics
 follow the chi-square distribution under the null hypothesis.
 For those that do not follow the chi-square distribution, the permutation procedure
 is often performed to estimate the p-value and power [Sha et al., 2007, Lin et al. 2009].

Previous attempts to find the asymptotic or approximate distribution of
 this class of statistics have been limited or case-specific.
 Tzeng et al. [2003] advanced our understanding of this area
 when they proposed a similarity-based statistic $T$
 and demonstrated that it approximately followed a normal distribution.
 The normal approximation works well under the null hypothesis provided that the sample
 sizes in the case and control populations are similar.
 However, the normal approximation can be inaccurate when the sample sizes differ,
 when there are rare haplotypes or when the alternative hypothesis is true instead,
 as we describe later.
 Schaid [2002] proposed the score test statistic to access the
 association between haplotypes and a wide variety of traits.
 Assuming normality of the response variables,
 this score test statistic can be written as a quadratic form of normal
 random variables and follows a chi-square distribution under the null hypothesis.
 To calculate power,
 Schaid [2005] discussed systematically how to find the non-central
 parameters under the alternative hypothesis.
 However, their result cannot be applied
 to the general case when a quadratic form statistic does not
 follow a non-central chi-square distribution.
 In the power comparisons made by Lin and Schaid [2009], power and p-values were all
 estimated using permutation procedures.
 However, a permutation procedure is usually
 not appropriate when the goal is to estimate a probability close to 0 or 1.
 Thus, if the true probability $p$ is about 0.01, 1,600 permutations are needed to derive
 an estimate that is between  $p/2$ and $3p/2$  with 95\% confidence.
 The number of permutations increases to 160,000 if $p$ is only 0.0001.
 Consequently, permutation tests are not suitable when a high level of significance is being sought.

The permutation procedure can also be very computationally intensive
 when estimating power.  In a typical power analysis, for example,
 the significance level is 0.05 and power is 0.8.
 Under these assumptions the p-value could be based on 1,000 permutations.
 Subsequently if the power of the test is estimated with 1,000 simulations,
 the statistic must be calculated 1,000,000 times.
 Moreover, to apply the multilocus association test method to genome-wide studies,
 the required significance level is many orders of magnitude below 0.05 to account for
 multiple comparisons and even 1,000 permutations will be completely inadequate.

Additional complications arise with permutations since most of the data in the
 current generation of association studies are un-phased genotypes.
 To explore the haplotype-trait association, the haplotypes are estimated
 using methods such as the EM-algorithm [Excoffier and Slatkin, 1995; Hawley and Kidd, 1995]
 or Bayesian procedures [Stephens and Donnelly, 2003].
 Two computational problems arise in this situation.
 First, the resulting haplotype distribution defines a very large
 category because all the haplotypes consistent with the corresponding genotypes will
 have a positive probability. Therefore, the number of rare haplotypes is usually greater
 than when phase is actually observed. Second, the process is again computationally
 intensive because the haplotype distribution needs to be determined for each permutation.
 To solve these problems, Sha et al. [2007] proposed a strategy where each rare haplotype
 is merged with its most similar common haplotype, thereby reducing the number of rare
 haplotypes and leading to a computationally efficient algorithm for the permutation procedure.
 This method is considerably faster than the standard EM algorithm. However,
 since it is still based on permutations it is not a perfect solution to the computational problem.
 Moreover, the process of  pruning out rare haplotypes can lead to systematic bias in the
 estimation of haplotype frequencies in some situations.

Based on these considerations, it is apparent that a fast and accurate way to estimate
 the corresponding p-value and associated power would be an important methodological step
 forward and make it possible to generalize the applications of these statistics.
 In this paper, we explore the asymptotic and approximate distribution of statistics with
 quadratic forms. Based on the results of these analyses, p-values and power can be
 estimated directly, eliminating the need for permutations.
 We assess the robustness of our methods using extensive simulation studies.

To simplify the notation, we use the statistic $S$ proposed by Sha et al. [2007] as an
 illustrative way to display our methods.
 We first assume that the similarity matrix $A$ is positive definite.
 We then extend this analysis to the case when $A$ is positive semi-definite and the more general case
 assuming symmetry of $A$ only. In the simulation studies, we use qq-plots and
 distances between distributions to explore the performance of our approximate distributions.
 In addition, we examine the accuracy of our approximations at the tails.
 Likewise, we assess the performance of our approximation under
 the alternative hypothesis by examining the qq-plots, distances, and tail probabilities.
 As an additional example, we apply our method to the statistic $T$ proposed by Tzeng et al.
 [2003] and compare the result with the normal approximation. Finally, we use our method
 to find the sample size needed for a candidate gene association study
 when linkage phase is unknown.

\noindent {\bf\Large Methods}

\noindent Assume that there are $k$ distinct
 haplotypes $(h_1,\cdots,h_k)$ with frequencies
 $p=(p_1,\cdots,p_k)^T$ in population 1, and $q=(q_1,\cdots,q_k)^T$
 in populations 2. To compare $p$ and $q$, we assume that sample 1 and sample 2
 are independent and are collected randomly from population 1 and population 2
 respectively. Let $n_j$ and $m_j$, $j = 1,\cdots,k$, represent
 the observed count of haplotype $h_j$ in sample 1 and sample 2
 respectively. We use the same notation as in Sha et al. [2007]:

\begin{itemize}
\item[] $n = \sum_{i=1}^{k} n_i$ = size of sample 1,
\item[] $m = \sum_{i=1}^{k} m_i$ = size of sample 2,
\item[] $\hat p = (\hat p_1, \cdots, \hat p_k)^T =
(n_1,\cdots,n_k)^T /n$,
\item[] $\hat q = (\hat q_1, \cdots, \hat q_k)^T =
(m_1,\cdots,m_k)^T /m$,
\item[] $a_{ij} = S(h_i,h_j)$ is the similarity score of haplotypes
$h_i$ and $h_j$,
\item[] $A = (a_{ij})$ is a $k\times k$ similarity matrix.
\end{itemize}

Let $s=p-q$ and $\hat s = \hat p - \hat q$. Then Sha et al.'s statistic is defined
 as $S=(\hat s^T A \hat s)/\sigma_0$, where $\sigma_0$ is the
 standard deviation of $\hat s^T A \hat s$ under the null hypothesis.
 In this paper, we focus on the distribution of $D_s=\hat s^T A \hat s$
 since $\sigma_0$ is a constant.

\noindent
{\bf Write $D_s$ as a function of  independent normal random variables}

\noindent
Assume that the observed haplotypes in sample 1 are independent
 and identically distributed (i.i.d.),
 then the counts of haplotypes $(n_1,\cdots,n_k)$ follow
 the multinomial distribution with parameters $(n;p_1,\cdots,p_k)$. Therefore,
 $\mu_p = E(\hat p) = p$ and $\Sigma_p = \mbox{Var}(\hat p) = (P - p p^T)/n$,
 where $P=\mbox{diag}(p_1,\cdots,p_k)$ is a
 $k\times k$ diagonal matrix.
 According to multivariate central limit theorem, $\hat p$
 asymptotically follows a multivariate normal distribution with mean $\mu_p$
 and variance $\Sigma_p$ when $n$ is large.
 A similar conclusion can be applied
 to $\hat q$ if replacing $p$ with $q$, $P$ with $Q$ and $n$ with
 $m$. Assume that samples 1 and 2 are independent. Then
 we conclude that $\hat s$ is asymptotically normally distributed with mean
 vector  $s = p-q$ and variance $\Sigma_s = \Sigma_p +
 \Sigma_q$.

Let $r_\sigma$ denote the rank of $\Sigma_s$.
 Then $r_\sigma \leq k-1$ since $\hat s = (\hat s_1,\cdots,\hat s_k)^T$
 only has $k-1$ free components. If we assume $p_i+q_i > 0$
 for all $i=1,\cdots,k$, then $r_\sigma=k-1$.
 Since $\Sigma_s$ is symmetric and positive semi-definite,
 there exists a $k \times k$
 orthogonal matrix $U=(u_1,\cdots,u_k)$,
 and diagonal matrix $\Lambda=\mbox{diag}(\lambda_1,\cdots,
 \lambda_{r_\sigma},0,\cdots,0)$, such that
 $\Sigma_s = U \Lambda U^T$ and
 $\lambda_1 \geq \cdots \geq \lambda_{r_\sigma} > 0$.

Now define matrices $U_\sigma = (u_1,\cdots,u_{r_\sigma})$,
 $\Lambda_\sigma = \mbox{diag}(\lambda_1,\cdots,\lambda_{r_\sigma})$,
 and $B = U_\sigma (\Lambda_\sigma)^{1\over 2}$.
 Then $\Sigma_s = U_\sigma \Lambda_\sigma U_\sigma^T = BB^T$
 and there exists $r_\sigma$
 independent standard normal random variables $Z = (Z_1, \cdots,
 Z_{r_\sigma})$ such that $\hat s \approx BZ + s$ for sufficiently large $n$ and $m$. Then we have
 \begin{eqnarray}
 D_s & = & \hat s^T A \hat s \nonumber \\
 &\approx& (BZ + s)^TA(BZ+s) \nonumber \\
 & = & Z^T B^T A B Z + 2 s^T A B Z + s^T A s
 \label{completeformula}
 \end{eqnarray}
 We then write $W = B^T A B=(\Lambda_\sigma)^{1\over 2}U_\sigma^T A
 U_\sigma (\Lambda_\sigma)^{1\over 2}$.
 Since $W$ is a $r_\sigma \times r_\sigma$ symmetric matrix, there always exists
 a $r_\sigma \times r_\sigma$ orthogonal matrix $V$ and a diagonal matrix
 $\Omega=\mbox{diag}(\omega_1,\cdots,\omega_{r_\sigma})$
 such that $W = V\Omega V^T$,
 where $\omega_1 \geq \cdots \geq \omega_{r_\sigma}$ are
 eigenvalues of $W$.

\noindent
{\bf Asymptotic and approximate distributions of $D_s$ with the assumption $s=0$}

Now let us consider the asymptotic distribution of $D_s$ under the null
 hypothesis $H_0: p = q$. That is, $s = 0$.
 Let $D_0$ represent the test
 statistic under $H_0$. Then we have
 $D_0 \approx Z^T W Z = Z^T V \Omega V^T Z$.
 Let $Y=(Y_1,\cdots,Y_{r_\sigma})^T = V^T Z$.
 Then $Y \sim N(0,I_{r_\sigma})$,
 where $I_{r_\sigma}$ is the $r_\sigma \times r_\sigma$
 identity matrix, and
 \begin{eqnarray}
 D_0 \approx \sum_{i = 1}^{r_\sigma}\omega_i Y_i^2 \label{nulldist}
 \end{eqnarray}

\noindent
{\it Case I: The similarity matrix $A$ is positive semi-definite}

\noindent
Under these assumptions $W$ will also be positive semi-definite.
 That is, $\omega_1 \geq \cdots \geq \omega_{r_\sigma} \geq 0$. Then $D_0$
 follows a weighted chi-square distribution asymptotically. To calculate the
 corresponding p-values efficiently, we could use a
 chi-square distribution to approximate it.

According to Satorra and Bentler [1994], the distribution of
 the adjusted statistic $\beta D_0$
 can be approximated by a central chi-square with
 degrees of freedom $df_0$, where $\beta$ is
 the scaling parameter based on the idea of Satterthwaite et al. [1941].
 This method is referred as {\em 2-cum chi-square approximation}
 since the parameters $\beta$ and $df_0$ are obtained by comparing
 the first two cumulants of the weighted chi-square and the chi-square.
 Specifically, let
 $\hat W$ be a consistent estimator of $W$.
 Then
 $$\beta D_0 \sim \chi_{df_0}^2$$
 approximately, where $\beta=\mbox{tr}(\hat W)/\mbox{tr}(\hat W^2)$,
 $df_0 = (\mbox{tr}(\hat W))^2/\mbox{tr}(\hat W^2)$, and
 tr($\cdot$) is the trace of a matrix.
 Note that it is not necessary
 to estimate $W$ because
 $\mbox{tr}(\hat W) = \mbox{tr}(\hat B^T A \hat B) =
 \mbox{tr}(A \hat B \hat B^T) = \mbox{tr}(A \hat \Sigma_s) \label{chitr1a}$, and
 $\mbox{tr}(\hat W^2) = \mbox{tr}(\hat B^T A \hat B \hat B^T A\hat B) =
 \mbox{tr}(A\hat \Sigma_s A \hat \Sigma_s)$,
 where $\hat \Sigma_s$ is a
 consistent estimate of $\Sigma_s$.

 Assume that the observed value of $D_s$ is $\hat{d}_s$. Then the
 p-value can be estimated using the following formula
 \begin{eqnarray}
 \mbox{p-value} & = & P_{H_0} (D_0 \geq \hat{d}_s)
 \approx P\left( \chi^2_{df_0} \geq
    \beta \hat{d}_s\right) \label{chisq1}
 \end{eqnarray}
Alternatively, assume that the significance level is $\alpha$
 and the value $c_\alpha^*$ is the quantile such that
 $P(\chi^2_{df_0} \geq c_\alpha^*) = \alpha$. Then the critical
 value of $D_s$ to reject $H_0$ at level $\alpha$ is
\begin{eqnarray}
d_\alpha^* & \approx & c_\alpha^*/\beta
   \label{chicritical}
\end{eqnarray}
 The above formulas indicate that the degrees
 of freedom $df_0$ and the coefficient $\beta$ of the
 chi-square approximation can be calculated directly
 from the similarity matrix and the variance matrix - a major advantage
 of this method since matrix decomposition can be very
 slow and inaccurate when the matrix has high dimensionality.

\noindent
{\it Case II: The similarity matrix $A$ is NOT positive semi-definite}

\noindent
In the above chi-square approximation, we assume that the
 similarity matrix $A$ is positive semi-definite.
 However, many similarity matrices do not satisfy this condition.
 For example, consider the length measure of the first
 5 haplotypes in Gene1 (Table 1 in Sha et al. 2007].
 The similarity between two haplotypes is defined as the maximum length of
 a common consecutive subsequence.
 The eigenvalues of the similarity
 matrix $A$ are $(2.84,1.21, 0.60, 0.36, -0.015)$.
 Therefore, $A$ is not positive semi-definite.

In this case, formula (\ref{nulldist}) is still
 true though formulas (\ref{chisq1})-(\ref{chicritical}) do not necessarily hold.
 A simple solution to this general case is to use the Monte Carlo method
 to estimate the p-value by generating independent
 chi-square random variables with known or estimated $\omega_i$.
 More specifically: Assume that the observed value of statistic $D_0$ is $\hat d_0$.
 Run $N$ simulations. For each simulation
 $t$, $t = 1,\cdots,N$, generate $r_\sigma$ independent standard normal
 random variables $y_{t1},\cdots,y_{t r_\sigma}$. Then calculate $d_t^0 =
 \sum_{j=1}^{r_\sigma} \omega_j y_{tj}^2$. The p-value can be estimated
 using the proportion of $d_t^0$ that is greater than or equal to
 $\hat d_0$. This method is not as good as the one
 based on formula (\ref{chisq1}), which
 calculates the p-value directly although, compared to the permutation
 procedure, it is computationally much simpler and faster.

Alternatively the eigenvalues can be separated into positive and negative groups.
 With estimated $w_i$, the sum of the positive group can be approximated
 by a single chi-square random variable, and as can the negative group.
 The corresponding p-value based on the difference of two chi-square
 random variables may be estimated by the Monte Carlo method
 or the technique described in Appendix D, which is used in all of our
 simulation studies.

\noindent
{\bf Asymptotic and approximate distributions of $D_s$ without the assumption $s=0$}

In this section, we would like to find the asymptotic distribution of $D_s$
 provided $p$ and $q$ are known but not
 necessarily equal. This is a typical situation for power analysis.
 In this case, the values of
 $s = p-q$ and $\Sigma_s = \Sigma_p +\Sigma_q
 = (P-pp^T)/n + (Q-qq^T)/m$
 are both known.
 Note that since $\Sigma_s$ is singular, it is not correct to write
 $D_s=(Z+B^{-1} s)^T B^T A B (Z+ B^{-1} s)$ since $B^{-1}$ is not
 well defined.
 Though $B^{-1}$ can be defined as the general inverse of $B$,
 it is impossible to find a $B^{-1}$ such that
 $B B^{-1} = I_k$ since its rank is at most $k-1$.
 Therefore, the following discussion for the case when $\Sigma_s$
 is singular is not as straightforward as that when
 $\Sigma_s$ is non-singular.

\noindent
{\it Case I: The similarity matrix $A$ is nonsingular}

\noindent
Then $W=B^T A B=(\Lambda_\sigma)^{1\over 2}U_\sigma^T A
 U_\sigma (\Lambda_\sigma)^{1\over 2}$ is nonsingular
 since $\Lambda_\sigma$ is nonsingular and
 $\mbox{rank}(U_\sigma)=r_\sigma$. So the eigenvalues
 of $W$ are non-zero. That is,
 $\omega_1 \neq 0, \cdots, \omega_{r_\sigma}\neq 0$.
 Therefore, $\Omega^{-1}=\mbox{diag}
 (1/\omega_1,\cdots,1/\omega_{r_\sigma})$ is well-defined.
 Let
 \begin{eqnarray}
 b & = & \Omega^{-1} V^T
    (\Lambda_\sigma)^{1\over 2} U_\sigma^T A s \nonumber\\
 c & = & s^T A s - b^T\Omega b \label{alternativepara},
 \end{eqnarray}
 Starting from equation (\ref{completeformula}),
 the statistic $D_s$ can be written as
 (see Appendix A for proof)
 \begin{eqnarray}
 D_s& = & (Y+b)^T \Omega (Y+b) +c
  = \sum_{i=1}^{r_\sigma}\omega_i (Y_i+b_i)^2 + c
  \label{alternativedist}
 \end{eqnarray}
 where $Y$ follows the multivariate standard normal distribution.
 Provided that the similarity matrix $A$ is positive definite,
 then $W$ will also be positive definite. We may assume that
 $\omega_1 \geq \cdots \geq \omega_{r_\sigma}> 0$.
 In this case, a non-central shifted chi-square distribution
 can be used for approximation.
 Note that when $\Sigma_s$ is non-singular, it is a special
 case of formula (\ref{alternativedist}) with
 $r_\sigma=k$, $U_\sigma=U$, and $\Lambda_\sigma=\Lambda$.
 In this case, it is easy to verify that $c=s^T A s - b^T \Omega b = 0$.

Liu et al. [2009] proposed a non-central shifted chi-square approximation
 for quadratic form $D=X^TAX$ by fitting the first four cumulants of $D$,
 where $A$ is positive semi-definite.
 In their settings, $X$ follows a multivariate normal distribution with a non-singular
 variance matrix. However, in our case, the rank of the variance matrix $\Sigma_s$
 is at most $k-1$.  Following the idea of Liu et al. [2009], we are able to
 derive the corresponding formula to fit our case
 (see Appendix B for details).
 Here we only define the necessary notation and
 list the final formula.  This method is
 referred as a {\em 4-cum chi-square approximation}.

Following Liu et al. [2009], define $\kappa_\nu= 2^{\nu-1} (\nu - 1)!
 (\mbox{tr}((A\Sigma_s)^\nu) + \nu s^T (A\Sigma_s)^{\nu-1} A s)$,
 $\nu=1,2,3,4$.
 Then let $s_1 = \kappa_3^2/(8\kappa_2^3)$ and $s_2 = \kappa_4/(12 \kappa_2^2)$.
 If $s_1 \leq s_2$, let $\delta=0$ and $df_a=1/s_1$.
 Otherwise, define $\xi = 1/(\sqrt{s}_1-\sqrt{s_1-s_2})$, and
 let $\delta=\xi^2(\xi\sqrt{s_1} - 1)$ and $df_a = \xi^2(3 - 2\xi \sqrt{s_1})$.
 Now let $\beta_1 = \sqrt{2(df_a+2\delta)/\kappa_{2}}$, and
 $\beta_2=df_a+\delta-\beta_1\kappa_1$. Then
 $$\beta_1 D_s + \beta_2 \sim \chi_{df_a}^2 (\delta)$$
 Let $d_\alpha^*$
 be the critical value as defined in equation (\ref{chicritical}).
 Then the power to reject $H_0$ at significance level $\alpha$
 can be estimated using the following formula:
 \begin{eqnarray}
 \mbox{power} & = & P_{H_a} (D_s \geq d_\alpha^*)
    \nonumber\\
 & \approx & P\left( \chi_{df_a}^2(\delta)\geq \beta_1 d_\alpha^* + \beta_2\right)
    \label{chisq2}
 \end{eqnarray}

Note that this 4-cum approximation is applicable not only
 under $H_a$, but also under $H_0$. Therefore, it can be used
 to find the p-value or define the critical value for rejection.
 Under $H_0$, the true haplotype frequencies $p$  and $q$  are usually
 unknown, although the difference $s=p-q$  is assumed to be zero.
 Therefore, to find the corresponding $\beta_1$
 and $\beta_2$, we can use $0$ to replace $s$ and
 $\hat \Sigma_s$ to replace $\Sigma_s$. Then the p-value is
 estimated as
 \begin{eqnarray}
 \mbox{p-value} & = & P_{H_0} (D_s \geq \hat d_s)
    \nonumber\\
 & \approx & P\left( \chi_{df_a}^2(\delta)\geq \beta_1 \hat d_s + \beta_2\right)
    \label{chisq2p}
 \end{eqnarray}
or alternatively, the critical value for rejection is
$$d_\alpha^* \approx (c_\alpha^* - \beta_2)/\beta_1,$$
where $c_\alpha^*$ is the quantile such that
$P(\chi_{df_a}^2 (\delta) \geq c_\alpha^*) = \alpha$.
Note that $\delta$ is automatically 0 if $s=0$.
To prove this, it is sufficient to prove that $s_1 \leq s_2$,
which is equivalent to
$[\mbox{tr}((A\Sigma_s)^3)]^2 \leq [\mbox{tr}((A\Sigma_s)^2)][\mbox{tr}((A\Sigma_s)^4)]$,
which is a direct conclusion from Yang et al. 2001.

If $A$ has negative eigenvalues, the approximations in
 formula (\ref{chisq2}) and (\ref{chisq2p})
 are not valid. However, equation (\ref{alternativedist}) is still true.
 In this case, we can use the same strategy as discussed
 in the case assuming $s=0$ to estimate the power or p-value.

\noindent
{\it Case II: The similarity matrix $A$ is singular}

\noindent
If $A$ is singular,
 that is, $\mbox{rank}(A)=r_a < k$, there exists
 an orthogonal matrix $G=(g_1,\cdots,g_k)$ and a diagonal matrix
 $\Gamma=\mbox{diag}(\gamma_1,\cdots,\gamma_{r_a},0,\cdots,0)$, where
 $\gamma_1 \neq 0, \cdots, \gamma_{r_a} \neq 0$, such that $A = G
 \Gamma G^T$. Let $G_a=(g_1,\cdots,g_{r_a})$ and
 $\Gamma_a=\mbox{diag}(\gamma_1,\cdots,\gamma_{r_a})$. Then $A$
 can be written as $A=G_a \Gamma_a G_a^T$.
 Now define $\hat s_a = G_a^T \hat s$. We have
 $D_s = \hat s^T A \hat s = \hat s_a^T \Gamma_a \hat s_a$, where
 $\Gamma_a$ is nonsingular and $\hat s_a$ asymptotically follows
 a normal distribution with mean $\mu_{a} = G_a^T s$ and variance
 $\Sigma_{a} = G_a^T \Sigma_s G_a$.
 Therefore, even if $A$ is singular, we can perform the above
 calculation to reduce its dimensionality and convert
 it into a non-singular matrix $\Gamma_a$.
 Then by replacing $s$ with $\mu_a$,
 $\Sigma_s$ with $\Sigma_a$,
 and $A$ with $\Gamma_a$, the discussion presented in {\it Case I}
 applies.

\noindent
{\bf Applications and extensions of our method}

\noindent
For illustrative purposes, we start the discussion with
 the statistic $D_s$ proposed by Sha et al (2007). Actually,
 our method can be applied to a much more general
 statistic $D$, as long as it can be written as the
 quadratic form
 $D=X^T A X$ with $X\sim N_k(\mu_x, \Sigma_x)$
 and $A$ being a $k\times k$ symmetric matrix which is not necessarily positive
 semi-definite.

When $\Sigma_x$ is nonsingular, the distribution of $D$ is
 straightforward because $D$ can be written as $D=(Z+b)^T
 \Sigma_x^{1\over 2}A\Sigma_x^{1\over 2} (Z+b)$, where $Z=(Z_1,\ldots,Z_k)^T$
 are i.i.d. normal random variables, and $b=\Sigma_x^{-\frac{1}{2}}\mu_x$
 with $\Sigma_x^{\frac{1}{2}}$ being a symmetric matrix with
 $\Sigma_x^{\frac{1}{2}} \Sigma_x^{\frac{1}{2}} = \Sigma_x$.
 Then $D$ follows a weighted
 non-central chi-square distribution. Moreover, if
 $\Sigma_x^{1\over 2}A\Sigma_x^{1\over 2}$ is idempotent, all the
 weights will be either 1 or 0. Therefore $D$ will follow a
 non-central chi-square distribution with degrees of freedom equal to
 the rank of $A$. However, when $\Sigma_x$ is singular,
 the above conclusion does not hold.
 In this paper, we not only show why $D$ can be written as a linear
 combination of chi-square random variables and how
 to estimate the corresponding parameter values, but also how
 to approximate its distribution using a chi-square or the difference of two chi-squares.
 To further illustrate the application of our method,
 we will discuss two more examples as follows.

First, let us consider the test statistic defined by Tzeng et al.
 (2003]. To keep the notation consistent with ours, the form of
 the statistic is written as $T = D_t/\sigma_0$, where
 $D_t = \hat p^T A \hat p - \hat q^T A \hat q$
 and $\sigma_0$ is the standard deviation of $D_t$ under the null hypothesis.
 It was claimed that
 $T$ is approximately distributed as a standard normal under the null
 hypothesis. However, we found that the normal approximation
 can be inappropriate in some situations.
 Write $D_p = \hat p^T A \hat p$ and $D_q =
 \hat q^T A \hat q$
 and assume that $A$ is positive definite.
 Then from our previous discussion, $D_p$ and
 $D_q$ both asymptotically follow a WNS-chi distribution when sample
 sizes $n$ and $m$ are large. However, their convergence rates
 differ when $n$ and $m$ are different.
 Then the normal approximation can be inaccurate when
 $n$ and $m$ are not very large.
 In fact, a difference in convergence rates is the same reason that
 the normal approximation is not applicable under the alternative hypothesis.
 We demonstrate this with simulation studies in the Results section.

Next, let us consider the statistic $S$ proposed by Schaid et al. [2002],
 where $S=(Y-\bar Y)^T X [(X-\bar X)^T (X-\bar X)]^{-1} X^T (Y-\bar
 Y)/\sigma_Y^2$ is defined based on the linear model $Y = \beta_0 +
 X\beta+\sigma_Y \varepsilon$ with $Y$ being the observed trait
 value, $X$ being the design matrix, $\beta =
 (\beta_1,\cdots,\beta_{k-1})$, and $\varepsilon$ being i.i.d.
 normal. Schaid [2005] assumed
 that $S$ follows a non-central chi-square
 distribution under the alternative hypothesis.
 Then the paper focused on the calculation of the
 non-central parameters under different situations of $X$
 (genotype, haplotype, or diplotype) and $Y$ (continuous or
 case-control). In fact, $S$ can be written as $S =
 (Y/\sigma_Y)^T A (Y/\sigma_Y)$, where
 $A = (X-\bar X)[(X-\bar X)^T (X-\bar X)]^{-1} (X-\bar X)^T $.
 Since $A^2 = A$, we
 conclude that $S$ follows a chi-square distribution with center $c
 = \mu_Y^T A \mu_Y^T/\sigma_Y^2$. In practice, $c$ can be replaced
 by its consistent estimate.

\noindent {\bf\Large Software}

We have integrated our approaches in an R source file quadrtic.approx.R.
 Given the mean $\mu_x$ and variance $\Sigma_x$ of $X$,
 this R file contains the subroutines to estimate
 (1) the probability $p = P\{X^T A X \leq d\}$ for a specific $d$,
 which is useful in approximating p-values or power;
 (2) the quantile $d^*$ such that $\alpha = P\{X^T A X \leq d^*\}$
 for a specific $\alpha$;
 and (3) the required sample size
 for a specific level of significance $\alpha$ and power $\beta$.
 This R file, as well as the readme and data files,
 can be downloaded from http://webpages.math.luc.edu/~ltong/software/.

\noindent {\bf\Large Results}

In the simulation studies we use the same four data sets
 as Sha et al. [2007]: Gene I, Gene II, Data I  and Data II
 (Tables I, IV and V in Sha et al. 2007], and the same
 three similarity measures:
 (1) the matching measure - score 1 for complete match and 0 otherwise;
 (2) the length measure - length spanned by the longest
 continuous interval of matching alleles; and
 (3) the counting measure - the proportion of alleles in common.
 We also explore the performance of our approximations
 using seven different sample sizes:
 $n=m=(20,50,100,500,1000,5000,10000)$.

\noindent {\bf Simulation studies based on the test statistic $D_s$}

We examine the performance of our approximations
 under both the null and the alternative hypotheses.

\noindent {\it Examining the distribution of $D_s$ under the null hypothesis}

Under the null hypothesis, we first examine the
 qq-plots of our 2-cum and 4-cum approximations
 for moderate sample size: $n=m=100$ (Figure 1).
 The $x$-axes are the quantiles of $D_s$,
 which are estimated based on 1.6 million independent simulations
 according to the true parameter values.
 The $y$-axes are the theoretical quantiles of our approximations
 based on the true parameter values.
 The range of the quantiles is from $0.00001$ to $0.99999$.
 For data 1 and data 2, the frequencies in the control population are used.
 From Figure 1, we observe that most of the points are around
 the straight line $y=x$, which leads to the conclusion
 that both the 2-cum and 4-cum approximations are very good in general,
 and even when there are rare haplotypes (gene 2, for example)
 and the sample size is moderate ($n=m=100$).
 Notice that at the left tails of these plots, the 4-cum approximation
 goes above the straight line $y=x$. However,
 this does not affect the performance of our
 approximations for p-values since only the right tail is of interest.
 At the right tails, the 2-cum approximations are all below
 the straight line, which indicates that the 2-cum approximation tends
 to under estimate the p-values. This is further verified in Table 2 below.
 The 4-cum approximation appears to perform better than the 2-cum.
 We also checked the qq-plots as the sample size increased.
 As expected, our approximations become better with larger sample sizes
 (results not shown here).

[Figure 1 about here]

The qq-plot can only show the comparison illustratively.
 However, it is also necessary to assess
 our approximations quantitatively. In this paper,
 we chose the two natural distances between any two distribution functions:
 the Kolmogorov distance (K-dist) and
 the Craimer-von Mises distance (CM-dist).
 For more distance choices, see Kohl and Ruckdeschel [2009].
 In general,
 the Kolmogorov distance measures the maximum differences between two
 distribution functions, while the Craimer-von Mises distance measures the
 average differences throughout the support of $x$ (See Appendix C for
 more details).
 We calculate the K-dist and CM-dist
 between our approximate distributions
 and the empirical ones
 based on 10K simulations under the null hypothesis
 for each combination of data set (4 in total),
 measure (3 in total) and sample size (7 in total).
 Notice that we did not use 1.6 million simulations here
 because it is computationally too intensive, especially when
 the sample size is large.
 In practice, we do not know the true values of $p$ and $q$.
 Therefore, the variance matrix $\Sigma_s$ is replaced by
 a consistent estimate $\hat \Sigma_s$, which will affect
 the accuracy of our approximations more or less.
 To account for the uncertainty when using $\hat \Sigma_s$,
 we simulate 20 samples and obtain an approximate distribution
 for each sample.

We compare the performance of the 2-cum approximation,
 the 4-cum approximation and the permutation procedure
 for different choices of sample sizes.
 We first use the true parameter
 values $p(=q)$ for the approximations (Table 1, rows ``true").
 Then we simulate 20 independent
 samples and replace $p(=q)$ and $\Sigma_s$ with
 $\hat \rho$ and $\hat \Sigma_s (\hat\rho)$
 (see Appendix E for definitions) respectively.
 The empirical distribution
 based on 1,000 permutations is also calculated
 for each of the 20 samples. Since the permutation procedure can be very slow
 when the sample sizes $n$ and $m$ are large, we did not perform permutations when
 $n=m\geq 1000$.
 For each method, the mean and standard deviation of distances based
 on these 20 samples are displayed in Table 1, rows ``mean" and ``s.d.".
 To simplify the output, we show only the results for Gene I using the matching measure.

 [Table 1 here]

From Table 1, we observe that for the 2-cum and 4-cum approximations,
 the mean distances using estimated parameter values converge to the distance
 using the true parameter values when
 sample size $n$ and $m$ increase. This is because both the asymptotic and the
 approximate components contribute to the distance. When sample sizes increase,
 the discrepancy due to the asymptotic component decreases eventually to zero,
 however, the discrepancy due to the approximate component does not.
 For example, the K-dist for the 4-cum method based on true parameter values
 decreases from 0.0630 to 0.0482 when the sample size increases from 20 to 50.
 But when the sample size increases from 50 to 10,000, it seems that
 this distance stays constant around 0.046.
 The 4-cum approximation appears better than 2-cum one if one cares about the average
 difference (CM-dist). Nevertheless, the opposite may be true when the maximum difference
 (K-dist) is preferred.  Compared with the permutation procedure, the proposed approximations
 show better performance for $n$ as small as 20, and comparable performance when $n$
 is reasonably large.  Note that our methods can be hundreds of times faster than
 permutations.

The conclusions regarding the convergence of the mean distances
 and the performance of permutations are similar when using
 the other data sets and measures. Therefore, in Table 2, we
 consider the distances based on true parameter values only.
 Moreover, since the main contributor to the distances is
 approximation when sample sizes are around 100, we use only
 the results from the case when $n=m=100$ in Table 2.

 [Table 2 about here]

From Table 2, we conclude that the 4-cum approximation performs
 better than the 2-cum approximation on average when sample sizes
 are moderate (around 100 individual haplotypes in each sample). However,
 there are some situations when the 2-cum approximation is preferred,
 such as those in the rows ``Gene1", ``DataII" and the column
 ``Counting" under ``K-dist" in Table 2.
 To find out how much of the distance is due to the discrete empirical
 distribution of $D_s$,
 we also checked the distance between the approximate distributions with
 their own empirical distributions based on 10K independent observations.
 The Kolmogorov distance is around 0.87\% and the Cramer-von Mises distance
 is around 0.38\%, which are about 20\% of the distances in Table 2.
 This indicates that when the predefined significance value is moderate, such as
 0.05, and the sample sizes are moderate, such as 100,
 both the 2-cum and the 4-cum approximations are appropriate.

In addition to its general performance,
 we would also like to know how good the approximations are when the significance
 level is very small. Ideally one should
 compare the approximations with true probabilities. However, since
 the theoretical distribution of $D_s$ is unknown, the only way to
 estimate the true probabilities is through simulations. When the true value
 of the probability is small, for example, $1\times 10^{-5}$, we need
 $1.6$ million simulations to ensure that the estimate is between
 $p/2$ and $3p/2$ with 95\% confidence.
 Here we consider moderate sample size $n=m=100$.
 We estimate the critical values for significance levels
 $\alpha=(0.05,0.01,0.001,0.0001,0.00001)$
 using the empirical distribution function of $D_s$ based on 1.6 million independent
 observations. For each combination of data set and similarity measure,
 we then estimate the corresponding significance levels using three methods:
 2-cum chi-square approximation, 4-cum chi-square approximation
 and a permutation procedure based on 160K million permutations.
 Since under the null hypothesis we need the sample proportions
 $\hat p$ and $\hat q$ for approximation, which will
 confound the effect of approximation with random errors,
 we examine the approximations based on both the true parameter values
 and the estimated ones from 20 simulations.
 It takes about 6 hours on a standard computer with
 Intel(R) Core(TM) CPU @ 2.66 GHz and 3.00 GB of RAM
 to estimate p-values using permutations for these four data sets,
 three measures and 20 simulations.
 However, only two seconds are needed using our approximations.
 Moreover, when the sample size increases,
 the computational time increases rapidly for a permutation procedure,
 while it stays the same for our approximations.

 [Tables 3 about here]

The results for Data II using the matching measures
 are summarized in Table 3.
 From this table, we can see that the 2-cum approximation
 performs slightly better than the 4-cum one when estimating
 a p-value around 0.05, while the 4-cum approximation is more accurate
 at p-values less than 0.01. This indicates that for a candidate
 gene study with significance level
 of 0.05, the 2-cum approximation
 is preferred since it is simpler and more accurate.
 However, for a genome screen, the 4-cum approximation would be more appropriate.
 Notice that the 4-cum approximation is accurate in estimation of a p-value
 as small as $0.1\%$. For probabilities around $0.01\%$,
 the 4-cum approximation tends to slightly under-estimate the true value and
 therefore will result in higher false positive results.
 For the probabilities around $0.001\%$, we list results in the last column
 of Table 3. However, since the number of simulations is limited,
 we can have only modest confidence in these approximations, although
 it is evident that they will provide an
 under-estimate of probabilities.
 Note that the permutation procedure gives good estimates for a p-value
 as small as 0.01\% due to large number of permutations.
 However, in the last column of Table 3, we notice
 that the standard deviation of estimated p-values is 0.001\%, which is
 about the same as the mean (0.0012\%) of these estimates.
 This is because 160K million permutations are far too few to give accurate
 estimate of a p-value of 0.001\%.
 The conclusions based on the other date sets are similar (results not shown).

\noindent {\it Examining the distribution of $D_s$ under the alternative hypothesis}

Similarly, we can examine the distribution of $D_s$ under the alternative
 hypothesis. For this purpose, we used Data 1 and 2 based on 160K simulations
 with sample sizes $n=m=100$.
 The range of the quantiles is from $0.0001$ to $0.9999$.
 Note that only the 4-cum approximation is available under the alternative hypothesis.
 From Figure 2, we observe that all the points lie close to
 a straight line, which indicates good approximations to the
 distribution of $D_s$ under the alternative hypothesis.

[Figure 2 about here]

Next, we examine the Kolmogorov and
 Cramer - von Mises distances between our approximations and
 the true distribution of $D_s$, which is estimated by
 the empirical distribution based on 10K simulations.
 The effect of sample size is similar to what was observed
 under the null hypothesis. So we consider only the case when
 $n=m=100$. Moreover, in this situation, we usually apply the formula
 to calculate power, in which case the true values of $p$ and $q$
 are assumed to be known.
 From Table 4, we notice that the distances are all less than 0.05. Therefore,
 it is safe to use the 4-cum approximation to find the power of $D_s$.

 [Table 4 about here]

Similarly, we examine the performance of the 4-cum approximation
 in the left tail, which is useful in a power analysis.
 In this situation, we assume that the parameter values are known.
 The quantiles at $(0.50, 0.60, 0.70, 0.80, 0.90, 0.95, 0.99)$
 are estimated through 160K simulations.
 Table 5 summarize the results when $n=m=20$, when $n=m=100$
 and when $n=m=1000$. From this table, we conclude that the
 power estimation is fairly accurate with moderate sample size
 ($n=m=100$) and moderate true power (less than 95\%).

 [Table 5 about here]

\noindent {\bf Simulations to check the distribution of the statistic $D_t$}

Tzeng et al. [2003] claimed that under the null hypothesis, the distribution of
 $D_t = \hat p^T A \hat p - \hat q^T A \hat q$ is approximately normal with
 mean 0 and variance $\mbox{Var}(D_t)$. This is true sometimes, but not
 always. In fact, if only the convergence rates of $\hat p^T A \hat p$
 and $\hat q^T A \hat q$ differ, the normal approximation will not be appropriate.
 This will occur under three situations. First, if there
 are several rare alleles, such as Gene 1 and Data 2, $\hat p$ and
 $\hat q$ can differ substantially even under null hypothesis (results not shown).
 Second, when the
 sample sizes $n$ and $m$ are not equal, the variances of $\hat p^T A \hat p$
 and $\hat q^T A \hat q$ will differ. Therefore, the convergence rates
 will differ (Figure 3). Third, under the alternative hypothesis, the convergence rates
 of $\hat p^T A \hat p$ and $\hat q^T A \hat q$ differ.
 Therefore, the normal approximation is not suitable for the above three situations.
 As an illustration, we use data set Data II and a matching measure to examine
 the qq-plot. The range of the quantiles is from 0.0001 to 0.9999. We first
 let $n=50$ and $m = 150$ and then let $n=1000$ and $m=3000$ (Figure 3).
 From figures 3, we can see that our 4-cum chi-square approximation
 can approximate the distribution of $D_t$ very well even when the smaller sample
 size is as small as 50. If the smaller sample size increases to 1000,
 the normal approximation also become acceptable.

 [Figures 3 about here]

To further compare the normal with the 4-cum chi-square approximation,
 we calculate the Kolmogorov and Cramer-von Mises distances for different combinations
 of data sets, measures and sample sizes. We assume that the size $m$ in
 the second sample is three times of the size $n$ in the first sample ($m=3n$).
 For illustration purpose, we show the results for Data II only (Table 6).
 From Table 6, we observe that the chi-square approximation has much smaller distances
 than the normal one, especially when sample sizes are not very large.
 The conclusions on the other data sets are similar.

 [Table 6 about here]

\noindent {\bf An example based on the estimation of power for a candidate gene study}

In this example we test the difference between haplotype distributions
 around the LCT gene (23 SNPs) found in populations HapMap3 CHB ($n = 160$) and HapMap3 JPT
 ($m = 164$). Since the linkage phase information is unknown, an EM algorithm was
 used to estimate the frequency of each distinct haplotype category.
 Under matching and length measures, the p-values of the test are both
 less than $10^{-8}$, which indicates a significant difference in haplotype
 distributions. However, these two similarity
 measures are very sensitive to errors due to genotyping or estimation
 and the results are therefore not reliable,
 especially in the case of unknown phase.
 Using a counting measure, the p-value is 0.026.
 It would then be interesting to know how many additional samples
 are required if we want power to be, say 90\%, at a significance level of 0.001,
 using the test statistic $D_s$ and the counting measure.
 Using the approximations described in our Methods section, we can easily
 calculate the required sample size.
 The quantities needed here are haplotype lists, frequencies and variance estimates
 for each population separately and jointly, which can be estimated using the
 EM algorithm. We first use the package haplo.stat [Sinnwell and Schaid, 2008]
 in R to find the starting value.
 Then we use a stochastic EM to refine the estimate and obtain the variance.
 The results are shown in Table 7. Note that all these calculations take only
 minutes on a standard computer with Intell(R) Core(TM) CPU @ 2.66 GHz and 3.00 GB
 of RAM. However, it requires at least several days to finish a single calculation using
 a permutation procedure.

 [Table 7 about here]

\noindent {\bf\Large Discussion}

In summary, the major contribution of the analytic approach presented in this paper
 is the description of the asymptotic and approximate distributions
 of a large class of quadratic form statistics used in multilocus association
 tests, as well as efficient ways to calculate the p-value and power of a test.
 Specifically, we have shown that the asymptotic distribution
 of the quadratic form $\hat s^T A \hat s$
 is a linear combination of chi-square distributions.
 In this situation, $\hat s$ asymptotically follows a
 multivariate normal distribution which may
 be degenerate.

To efficiently calculate the p-value under the null hypothesis $s=E(\hat s)=0$,
 we propose 2-cum and 4-cum chi-square approximations to the distribution of $\hat s^TA\hat s$.
 We extended the 4-cum approximation in Liu et al. [2009] to allow degenerate $\hat s$
 and general symmetric $A$ which may not be positive semi-definite.
 Generally speaking, the 4-cum is better than the 2-cum approximation
 when dealing with probabilities less than 0.01. Nevertheless, the latter may perform
 better for moderate probabilities, say 0.05.  On the other hand, the 2-cum
 method only involves the products of up to two $k\times k$ matrices, while the 4-cum approach
 relies on a product of four $k\times k$ matrices. When the number of haplotypes $k$ is large,
 the 2-cum approach is computationally much less intensive.
 To estimate the power of a test, however, only the 4-cum approximation is valid.

The similarity matrix $A$ can be singular or approximately singular due to missing values.
 In this case, we decompose $A$ and perform
 dimension reduction to get a smaller but nonsingular similarity matrix.
 The most attractive feature of our method is that we do not need to decompose matrices $\Sigma_s$ or $W$
 when $A$ is positive semi-definite
 because the decompositions do not appear in the final formula.
 This not only simplifies the formula, but also results in
 better computational properties
 since it is often hard to estimate $\Sigma_s$ accurately.


In this paper we do not consider the effect of latent population structure.
 It has been widely recognized that the presence of undetected population structure
 can lead to a higher false positive error rate or to decreased power of association testing
 [Marchini et al. 2004]. Several statistical methods have been developed to adjust for
 population structure [ Devlin and Roeder 1999, Prichard and Rosenberg 1999, Pritchard et al. 2000,
 Reich and Goldstein 2001, Bacanu et al. 2002, Price et al. 2006].
 These methods mainly focus on the effect of population stratification on
 the Cochran-Armitage chi-square test statistic.
 It would be interesting to know how these methods can be applied to
 the similarity or distance-based statistic to conduct association
 studies in the presence of population structure.

Our methods can potentially be applied to the genome-wide association studies because
 the computations are fast and small probabilities can be estimated
 with acceptable variation.
 To perform a genome screen one must define the regions of interest manually,
 which will be exceedingly tedious.
 However, due to limitation in length, we do not discuss the problem of
 how to define haplotype regions automatically.
 Clearly before this approach can be applied in practice,
 such methods and software will have to be developed.
 We also propose to explore this issue in the future.

\noindent {\bf\Large Acknowledgements}

This work was supported in part by grants from the NHLBI (RO1HL053353) and
the Charles R. Bronfman Institute for Personalized Medicine at Mount Sinai Medical Center (NY).
We are grateful to suggestions by Dr. Mary Sara McPeek and students from her statistical genetics seminar class.

\noindent {\bf\Large Reference}

Bacanu S-A, Devlin B, Roeder K (2002) Association studies for quantitative traits in structured populations. Genet Epidemiol 22 (1): 78–93.

Bentler PM, Xie J (2000) Corrections to test statistics in principal Hessian directions. Statistics and Probability Letters 47: 381-389.

Devlin B, Roeder K (1999) Genomic control for association studies. Biometrics 55: 997-1004.

Driscoll MF (1999) An improved result relating quadratic forms and chi-square distributions. The American Statistician 53: 273-275.

Excoffier L, Slatkin M (1995) Maximum likelihood estimation of molecular haplotype frequencies in a diploid population. Mol Biol Evol 12: 921-927.

Hawley M, Kidd K (1995) Haplo: a program using the EM algorithm to estimate the frequencies of multi-site haplotypes. J Hered 86: 409-411.

Kohl M, Ruckdeschel P (2009) The distrEx Package, available via\\ http://cran.r- project.org/web/packages/distrEx/distrEx.pdf

Liu H, Tang Y, Zhang HH (2009) A new chi-square approximation to the distribution of non-negative definite quadratic forms in non-central normal variables. Computational Statistics and Data Analysis 53: 853-856.

Lin WY, Schaid DJ (2009) Power comparisons between similarity-based multilocus association methods, logistic regression, and score tests for haplotypes. Genet Epidemiol 33 (3): 183-197.

Marchini J, Cardon LR, Phillips MS, Donnelly P (2004) The effects of human population structure on large genetic association studies. Nature Genetics 36, 512-517.

Marquard V, Beckmann L, Bermejo JL, Fischer C, Chang-Claude J (2007) Comparison of measures for haplotype similarity. BMC Proceedings 1 (Suppl 1): S128.

Price AL, Patterson NJ, Plenge RM, Weinblatt ME, Shadick NA, Reich D (2006) Principal components analysis corrects for stratification in genome-wide association. Nature Genetics 38:904-909

Pritchard JK, Rosenberg NA (1999) Use of unlinked genetic markers to detect population stratification in association studies. American Journal of Human Genetics 65:220-228.

Pritchard JK, Stephens M, Rosenberg NA, Donnelly P (2000) Association mapping in structured populations. American Journal of Human Genetics 67: 170-181

Reich DE, Goldstein DB (2001) Detecting association in a case-control study while correcting for population stratification. Genet Epidemiol 20 (1): 4–16.

Schaid DJ, Rowland CM, Tines DE, Jacobson RM, Poland GA (2002) Score tests for association between traits and haplotypes when linkage phase is ambiguous. Am. J. Hum. Genet. 70: 425-434.

Schaid DJ (2005) Power and sample size for testing associations of haplotypes with complex traits. Annals of Human Genetics 70: 116-130.

Sha Q, Chen HS, Zhang S (2007) A new association test using haploltype similarity. Genetic Epidemiology 31: 577-593.

Sinnwell JP, Schaid DJ (2008). http://mayoresearch.mayo.edu/mayo/research/schaid\_lab/\\software.cfm

Solomon H, Stephens MA (1977) Distribution of a sum of weighted chi-square variables. Journal of the American Statistical Association 72: 881-885.

Stephens M, Donnelly P (2003)  A comparison of Bayesian methods for haplotype reconstruction from population genotype data. Am. J. Hum. Genet. 73:1162-1169.

Tzeng JY, Devlin B, Wasserman L, Roeder K (2003) On the identification of disease mutations by the analysis of haplotype similarity and goodness of fit. Am. J. Hum. Genet. 72: 891-902.

Tzeng JY, Zhang D (2007) Haploltype-based association analysis via variance-components score test. Am. J. Hum. Genet. 81: 927-938.

Yang XM, Yang XQ, Teo KL (2001) A Matrix Trace Inequality. Journal of Mathematical Analysis and Applications 263: 327–331.

\noindent {\bf\Large Appendix}

\noindent
{\bf A: Proof that $D_s$ can be written as a linear combination of independent chi-square random variables
 under the alternative hypothesis}

\noindent
Start with (\ref{completeformula}) and
 $W = B^T A B=V\Omega V^T$. Then
 $$Z^TB^TABZ=Z^TWZ=Z^TV\cdot\Omega\cdot V^TZ=Y^T\Omega Y$$
 $$s^TABZ=s^TABV\Omega^{-1}\cdot\Omega\cdot V^TZ=b^T\Omega Y$$
 where $Y=V^TZ\sim N(0,I_{r_\sigma})$.
 Let $c=s^TAs-b^T\Omega b$. We have
 \begin{eqnarray*}
 D_s &\approx& Z^TB^TABZ+2s^TABZ+s^TAs\\
 &=&Y^T\Omega Y+2b^T\Omega Y+s^TAs\\
 &=&(Y+b)^T\Omega(Y+b)+s^TAs-b^T\Omega b\\
 &=& \sum_{i=1}^{r_\sigma} \omega_i (Y_i + b_i)^2 + c
 \end{eqnarray*}

\noindent
{\bf B: Four-cumulant non-central chi-square approximation}

\noindent
Rewrite the original statistic $D_s=\hat s^TA\hat s$ into
 its asymptotic form $(Y+b)^T\Omega(Y+b)+c$ (see Appendix A).
 We only need to consider the shifted quadratic form $$Q(Y_b)=Y_b^T\Omega Y_b+c$$
 (see (\ref{alternativedist})), where $Y_b = Y+b \sim N(b, I_{r_\sigma})$, and
 $\Omega={\rm diag}(\omega_1, \ldots, \omega_{r_\sigma})$ with
 $\omega_1\geq\omega_2\geq\cdots\geq \omega_{r_\sigma}>0$.

According to Liu et al. [2009], the $\nu$th cumulant of $Q(Y_b)$ is
 $$\kappa_\nu=2^{\nu-1}(\nu-1)!(\kappa_{\nu,1}+\nu \kappa_{\nu,2})$$
 In our case, for $\nu=1,2,3,4$,
 $$\kappa_{\nu,1}={\rm tr}(\Omega^\nu)={\rm tr}((V^TWV)^\nu)
 ={\rm tr}(W^\nu)={\rm tr}((B^TAB)^\nu)={\rm tr}((A\Sigma_s)^\nu)$$
 And for $\nu = 1$,
 $$\kappa_{\nu,2}=b^T\Omega b+c=b^T\Omega b+s^TAs-b^T\Omega b=s^TAs$$
 For $\nu=2,3,4$,
 \begin{eqnarray*}
 \kappa_{\nu,2} &=& b^T\Omega^\nu b\\
  &=& s^TAU_\sigma(\Lambda_\sigma)^{\frac{1}{2}}V\Omega^{-1}\cdot\Omega^\nu
      \cdot\Omega^{-1}V^T(\Lambda_\sigma)^{\frac{1}{2}}U_\sigma^TAs\\
  &=& s^TAU_\sigma(\Lambda_\sigma)^{\frac{1}{2}}V\Omega^{\nu-2}
      V^T(\Lambda_\sigma)^{\frac{1}{2}}U_\sigma^TAs\\
  &=& s^TAB(V\Omega V^T)^{\nu-2}B^TAs\\
  &=& s^TAB(B^TAB)^{\nu-2}B^TAs\\
  &=& s^T(A\Sigma_s)^{\nu-1}As
 \end{eqnarray*}
 Therefore,
 $$\kappa_\nu= 2^{\nu-1} (\nu - 1)! (\mbox{tr}((A\Sigma_s)^\nu)
   + \nu s^T (A\Sigma_s)^{\nu-1} A s), \ \nu=1,2,3,4$$
 which actually takes the same form as in Liu et al. [2009].
 So the discussion here extends Liu et Al. [2009]'s formulas
 to more general quadratic form which allows degenerate multivariate normal distribution.

\noindent
{\bf C: Distance between a continuous distribution and an empirical distribution}

To compare one continuous cumulative distribution
 function $F_1$ and one empirical distribution $F_2$ (or discrete distribution),
 two natural distances are the Kolmogorov distance
 $$d_K(F_1, F_2)=\sup_x \left|F_1(x)-F_2(x)\right|$$
 and the Cramer-von Mises distance with measure $\mu=F_1$
 $$d_{cv}(F_1,F_2)=\left(\int [F_1(x)-F_2(x)]^2 d F_1(x)\right)^{\frac{1}{2}}$$

Note that $F_2$ is piecewise constant.
 Let $x_1,x_2,\ldots,x_n$ be all distinct discontinuous points of $F_2$.
 We keep them in an increasing order.
 If $F_2$ is an empirical distribution, $x_1,x_2,\ldots,x_n$ are distinct
 values of the random sample which generates $F_2$.
 Write $x_0=-\infty$.

For Kolmogorov distance, the maximum
 can be obtained by checking all the discontinuous points of $F_2$.  Therefore,
 $$d_K(F_1, F_2)=\max_i\left\{|F_1(x_i)-F_2(x_i)|\right\}\bigvee
 \max_i\left\{|F_1(x_i)-F_2(x_{i-1})|\right\}$$

For Cramer-von Mises distance,
 \begin{eqnarray*}
 d^2_{cv}(F_1,F_2)
 &=&\int_{-\infty}^{x_1}F_1(x)^2dF_1(x)+\int_{x_n}^{\infty}[1-F_1(x)]^2dF_1(x)\\
 & &+\sum_{i=1}^{n-1}\int_{x_i}^{x_{i+1}}  [F_1(x)-F_2(x_i)]^2dF_1(x)\\
 &=& \frac{1}{3}F_1^3(x_1)+\frac{1}{3}[1-F_1(x_n)]^3\\
 & & +\frac{1}{3}\sum_{i=1}^{n-1}\left\{
 [F_1(x_{i+1})-F_2(x_i)]^3-[F_1(x_i)-F_2(x_i)]^3\right\}
 \end{eqnarray*}
 Note that the formulas above work better than the corresponding {\tt R} functions
 in the package "distrEx" (downloadable via http://cran.r-project.org/).
 Those {\tt R} functions have difficulties with large sample sizes (say $n\geq 2000$),
 because their calculation replies on the grids on the real line.

\noindent{\bf D: Calculate the difference between two non-central chi-squares}

Let $Y_1$ and $Y_2$ be two independent non-central chi-square random variables
with probability density function $f_1(y)$ and $f_2(y)$ respectively.
Write $Z=Y_1-Y_2$.  Then
the probability density function $f(z)$ of $Z$ can be calculated through
\begin{eqnarray*}
f(z) &=& \int_{-\infty}^{\infty}f_1(z+y)f_2(y)dy\\
&=& \int_0^1 f_1\left(z+\log\frac{x}{1-x}\right)f_2\left(\log\frac{x}{1-x}\right)
\cdot\frac{1}{x(1-x)}dx
\end{eqnarray*}

The cumulative distribution function $F(z)$ of $Z$ can be calculated through
\begin{eqnarray*}
F(z) &=& \int_{-\infty}^{\infty}\int_{-\infty}^z f_1(y_1+y_2) f_2(y_2) dy_1dy_2\\
&=& \int_0^1 \int_0^{\frac{e^z}{1+e^z}}
f_1\left(\log\frac{x_1x_2}{(1-x_1)(1-x_2)}\right)
f_2\left(\log\frac{x_2}{1-x_2}\right)\\
& &\cdot\frac{1}{x_1x_2(1-x_1)(1-x_2)}dx_1dx_2
\end{eqnarray*}
Note that we perform the transformation $y=\log\left(x/(1-x)\right)$ in both formulas to
convert the integrating interval from
$(-\infty,\infty)$ into $(0,1)$ for numerical integration purpose.

\noindent{\bf E: Simplified formulas for tr($\hat W$) and tr($\hat W^2$) when phase is known}

 Let $\hat \rho = (\hat \rho_1,
 \cdots, \hat \rho_k)$, where $\hat \rho_i = (n \hat p_i + m \hat q_i)/(n+m)$,
 $i=1,\ldots,k$. Then
 under the null hypothesis, $\hat \rho_i$ is a consistent estimate of
 $p_i$ ($=q_i$). It follows that $\hat \Sigma_s = \hat
 \Sigma_s(\hat \rho) = (1/n + 1/m)(\hat R - \hat \rho \hat \rho^T)$ is a
 consistent estimate of $\Sigma_s$, where
 $\hat R = \mbox{diag} (\hat \rho_1, \cdots, \hat \rho_k)$.
 Since $\hat R$ is a diagonal matrix and $\hat \rho$ is a vector,
 the calcualtion of tr($\hat W$) and tr($\hat W^2$) can be further simplified as
 \begin{eqnarray*}
 \mbox{tr}(\hat W)
 & = & \left({1\over n} +{1 \over m}\right)
    \left(\sum_{j=1}^k a_{jj}\hat \rho_j(1-\hat \rho_j)
    - 2\sum_{j_1=1}^k \sum_{j_2>j_1} a_{j_1 j_2}\hat \rho_{j_1} \hat
    \rho_{j_2}\right) \label{chitr1b} \\ \\
 \mbox{tr}(\hat W^2)
 & = & \left({1\over n}+{1\over m}\right)^2\left[
    \sum_{j=1}^k a_{jj}^2 (1-\hat \rho_j)^2 \right. \\
 && + \left. 2\sum_{j_1=1}^k \sum_{j_2 > j_1} a_{j_1 j_2}^2
    \hat \rho_{j_1} \hat \rho_{j_2} (1-\hat \rho_{j_1}-\hat \rho_{j_2}) \right.\\
 && \left. - 4\sum_{j_1=1}^k \sum_{j_2 > j_1} \hat \rho_{j_1} \hat \rho_{j_2}
   \sum_{l=1}^k a_{l j_1} a_{l j_2} \hat \rho_l \right. \\
 && \left. + \left(\sum_{j=1}^k a_{jj} \hat \rho_j^2 + 2 \sum_{j_1=1}^k
    \sum_{j_2 > j_1} a_{j_1 j_2} \hat \rho_{j_1} \hat \rho_{j_2}
    \right)^2\right]
 \end{eqnarray*}

 It is important to point out that the degrees of freedom
 $df_0=\mbox{tr}(\hat W)^2/\mbox{tr}(\hat W^2)$
 do not depend on sample sizes $n$ and $m$ according
 to the above formulas.

\newpage

\centerline{\bf\Large Figures}

\begin{center}
\begin{quote}
{\bf Figure 1:} The qq-plots of
 the 2-cum (red) and 4-cum (blue) approximations
 to the distribution of $D_s$ (based on 1.6 million simulations)
 under the null hypothesis using gene 1
 (first row), gene 2 (second row), data 1 (third row)
 and data 2 (fourth row). The black solid line is $y=x$.
 We use the true values of $p$ and $q$ here.
 The left, middle, and right
 columns are for matching, length, and counting measures respectively.
 The sample sizes are $m=n=100$.
\end{quote}
\hspace{0.0cm}
\psfig{figure=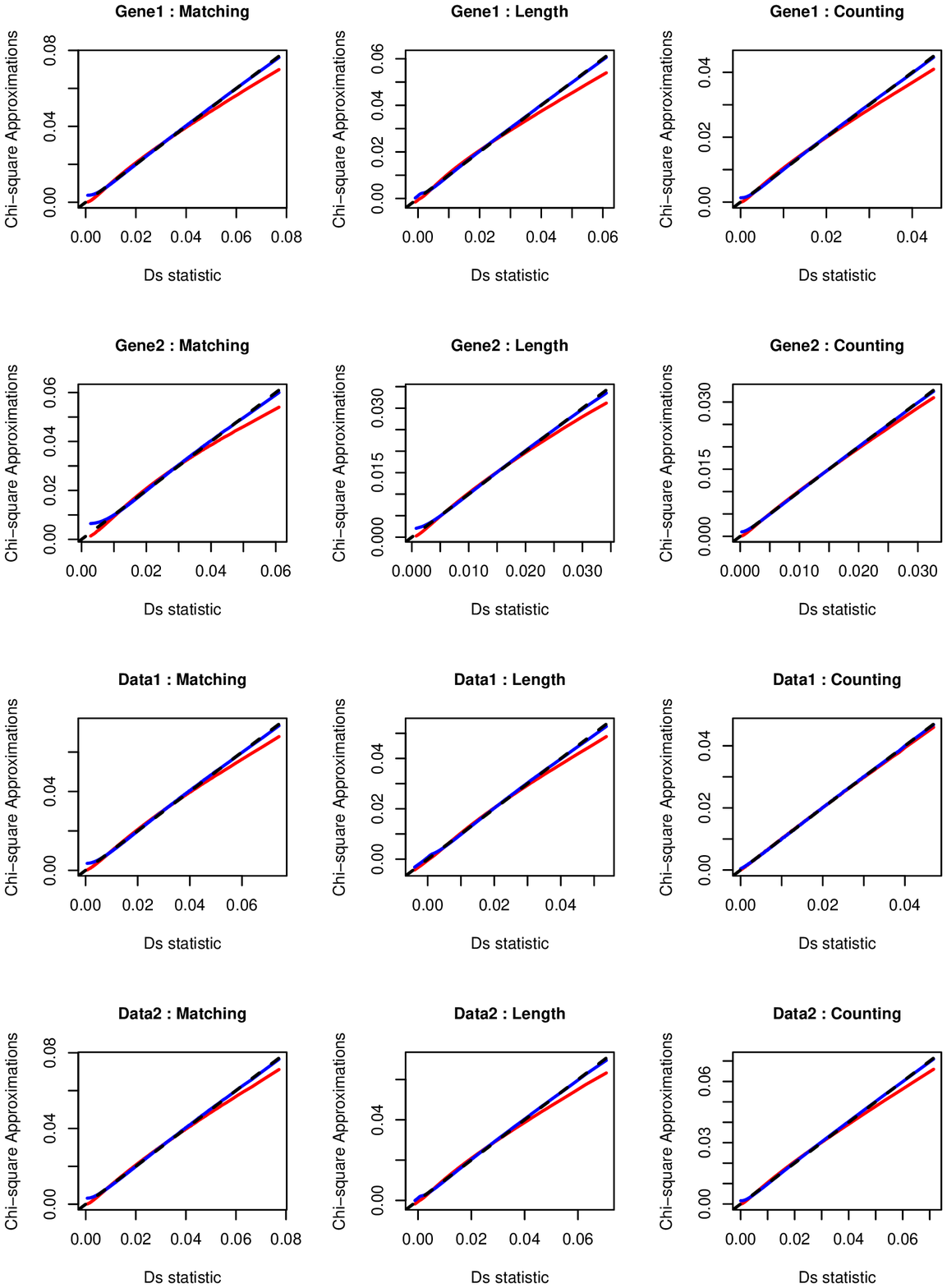,height=6.5in,width=5in,angle=0}
\end{center}

\begin{center}
\begin{quote}
{\bf Figure 2:} The qq-plots of the 4-cum (blue) approximations
to the distribution of $D_s$ (based on 160K simulations)
under the alternative hypothesis using data 1
(first row) and data 2 (second row).
The black solid line is $y=x$.
We use the true values of $p$ and $q$ here.
The left, middle, and right
columns are for matching, length, and counting measures respectively.
The sample sizes are $m=n=100$.
\end{quote}
\hspace{0.0cm}
\psfig{figure=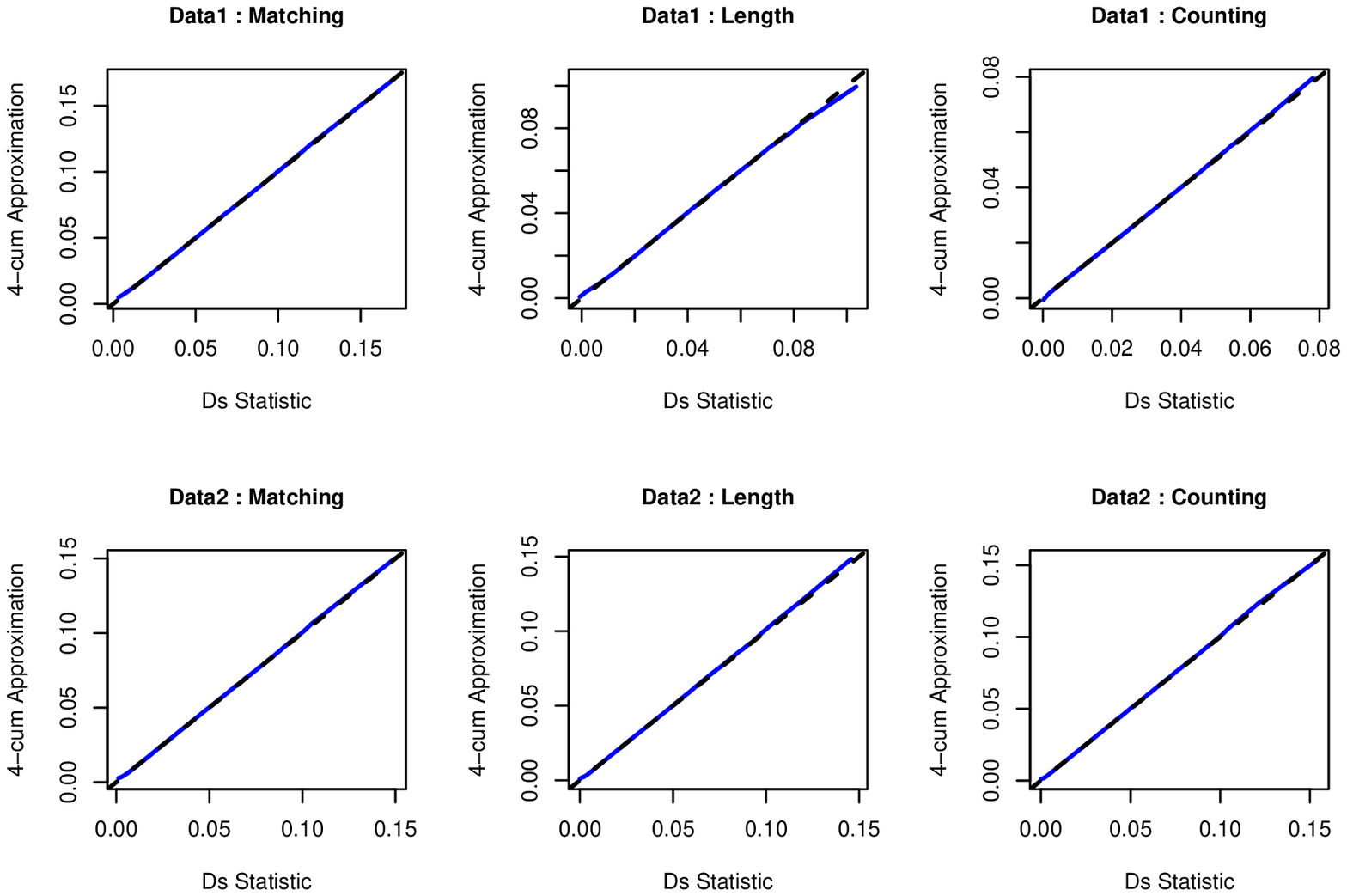,height=3.5in,width=5in,angle=0}
\end{center}

\newpage
\begin{center}
\begin{quote}
{\bf Figure 3:} The qq-plots of the 4-cum chi-square approximation
(blue ``4") and the normal approximation (red ``n")
to the distribution of $D_t$ under the null hypothesis using
Gene II and the matching measure.
We use the true values of $p$ and $q$ here.
The left plot has a smaller sample size $n=50$ and $m=150$.
The right plot has a larger sample size $n=1000$ and $m=3000$.
\end{quote}
\hspace{0.0cm}
\psfig{figure=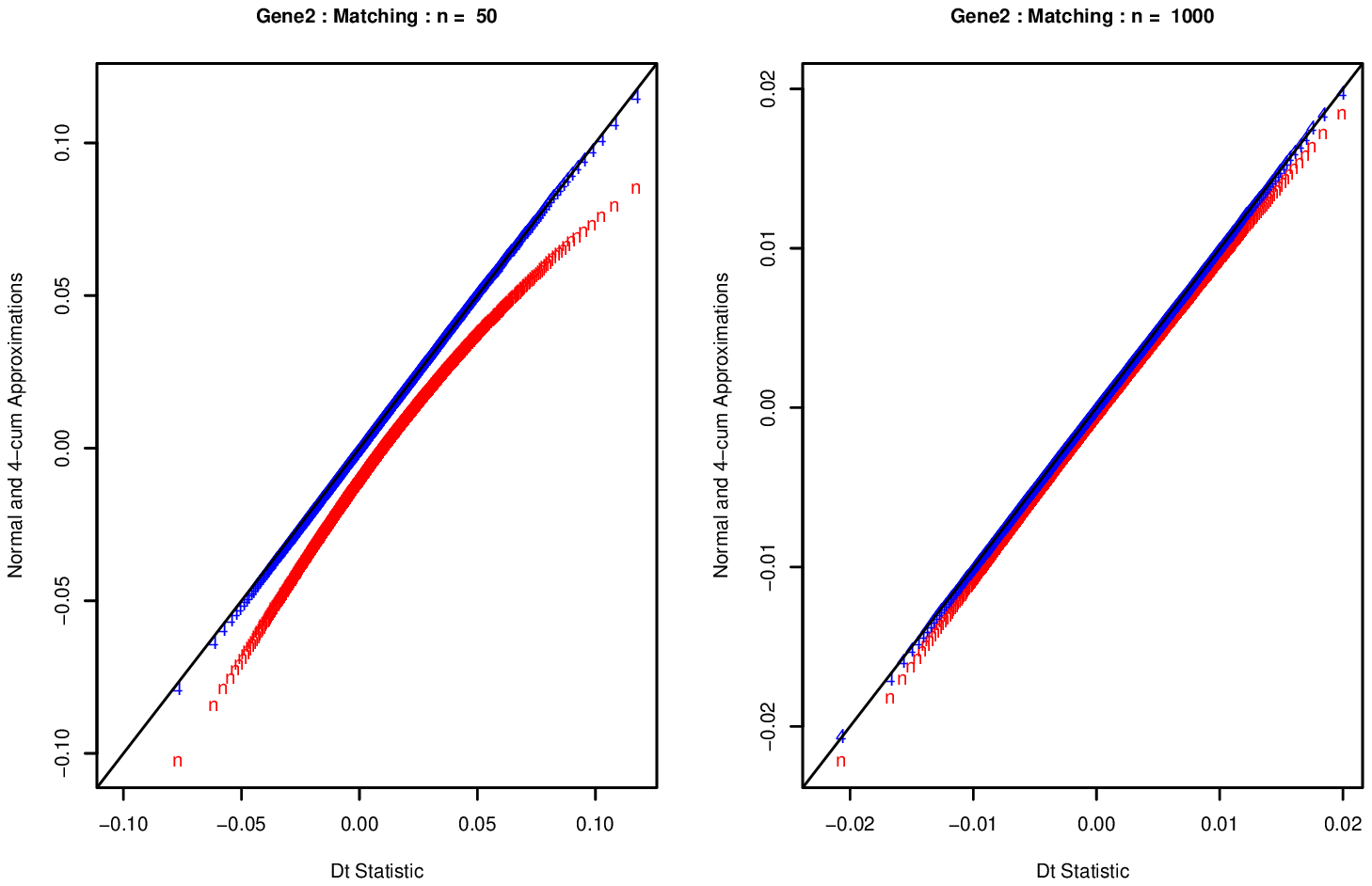,height=2.5in,width=5in,angle=0}
\end{center}

\newpage

\centerline {\bf\Large Tables}

\small\begin{center}\begin{tabular}{lllccccccc}
\multicolumn{10}{l}{\bf TABLE 1. Kolmogorov and Cramer-von Mises distances (\%) under the null}\\
\multicolumn{10}{l}{\bf hypothesis for Gene I using matching measure} \\\\ \hline\hline
        &      &       & \multicolumn{7}{c}{sample size ($n = m$)}\\ \cline{4-10}
Distance&Method&       &   20 &  50 & 100 & 500 &1000 &5000 &10000\\ \hline\hline
        &      &  true & 5.72 &4.95 &4.71 &4.13 &3.77 &4.23 &3.69\\
        & 2-cum&  mean & 8.69 &7.55 &5.68 &4.21 &4.00 &4.23 &3.70\\
        &      &  s.d. & 2.71 &2.90 &1.55 &0.56 &0.45 &0.21 &0.14\\ \cline{2-10}
        &      &  true & 6.30 &4.82 &4.54 &4.65 &4.70 &4.51 &4.75\\
K-dist  & 4-cum&  mean & 8.76 &6.81 &4.80 &4.57 &4.61 &4.52 &4.77\\
        &      &  s.d. & 3.43 &3.37 &1.11 &0.48 &0.34 &0.14 &0.09\\ \cline{2-10}
        & perm.&  mean &10.39 &6.74 &4.16 &3.00 &  NA &  NA &  NA\\
        &      &  s.d. & 3.16 &2.89 &1.15 &1.18 &  NA &  NA &  NA\\ \hline\hline
        &      &  true & 2.25 &2.35 &2.05 &2.21 &2.00 &2.31 &2.02\\
        & 2-cum&  mean & 4.18 &3.81 &2.63 &2.30 &2.08 &2.31 &2.02\\
        &      &  s.d. & 1.71 &1.67 &0.68 &0.11 &0.14 &0.04 &0.02\\ \cline{2-10}
        &      &  true & 1.98 &1.47 &1.24 &1.20 &1.38 &1.52 &1.31\\
CM-dist & 4-cum&  mean & 4.15 &3.38 &2.10 &1.35 &1.54 &1.53 &1.32\\
        &      &  s.d. & 2.23 &2.24 &1.03 &0.26 &0.23 &0.10 &0.05\\ \cline{2-10}
        & perm.&  mean & 4.32 &3.21 &1.96 &1.29 &  NA &  NA &  NA\\
        &      &  s.d. & 2.27 &1.91 &0.71 &0.70 &  NA &  NA &  NA\\ \hline\hline
\end{tabular}\end{center}
\normalsize

\small\begin{center}\begin{tabular}{ccccccccc}
\multicolumn{9}{l}{\bf TABLE 2. Kolmogorov and Cramer-von Mises distances under the null hypothesis}\\
\multicolumn{9}{l}{\bf when sample sizes $n=m=100$} \\ \hline\hline
       &      &\multicolumn{3}{c}{K-dist}& &\multicolumn{3}{c}{CM-dist}\\ \cline{3-5} \cline{7-9}
Data   &Method&Matching&Length &Counting & &Matching&Length &Counting\\ \hline\hline
Gene I &2-cum & 4.71& 7.89&  5.52& & 2.05& 3.73& 2.78  \\
       &4-cum & 4.54& 9.25& 10.50& & 1.24& 3.29& 2.84  \\ \hline\hline
Gene II&2-cum & 3.84& 2.57&  2.19& & 2.07& 1.55& 1.26  \\
       &4-cum & 2.85& 1.74&  1.45& & 1.21& 0.68& 0.61  \\ \hline\hline
Data I &2-cum & 3.12& 4.02&  1.59& & 1.59& 2.09& 0.69  \\
       &4-cum & 4.15& 3.97&  2.16& & 1.62& 1.48& 0.66  \\ \hline\hline
Data II&2-cum & 3.80& 6.43&  6.28& & 1.71& 3.17& 2.96  \\
       &4-cum & 3.92& 8.12& 10.99& & 1.08& 2.46& 2.73  \\ \hline\hline
\end{tabular}\end{center}
\normalsize

\small\begin{center}\begin{tabular}{lllllllll}
\multicolumn{9}{l}{\bf TABLE 3. Comparison of probabilities in the right tail for Data II using}\\
\multicolumn{9}{l}{\bf matching measure when $n=m=100$.}\\ \hline\hline
       &       &     &  \multicolumn{5}{c}{$ p = \% $}\\ \cline{4-9}
Data   & Method&     &  5     & 1     & 0.1   & 0.01  & 0.001\\ \hline\hline
       &       &true &  4.9724& 0.7977& 0.0483& 0.0024& 0.0002\\
       & 2-cum &mean &  5.0302& 0.8134& 0.0503& 0.0027& 0.0002\\
       &       &s.d. &  0.1619& 0.0733& 0.0102& 0.0009& 0.0001\\ \cline{2-9}
Data II&       &true &  5.1828& 1.0273& 0.0929& 0.0076& 0.0008\\
       & 4-cum &mean &  5.2266& 1.0297& 0.0926& 0.0076& 0.0008\\
       &       &s.d. &  0.1331& 0.0753& 0.0161& 0.0022& 0.0003\\ \cline{2-9}
       & perm. &mean &  5.0482& 0.9976& 0.1011& 0.0104& 0.0012\\
       &       &s.d. &  0.1602& 0.0771& 0.0238& 0.0033& 0.0010\\ \hline\hline
\end{tabular}\end{center}

\normalsize

\small\begin{center}\begin{tabular}{cccccccc}
\multicolumn{8}{l}{\bf TABLE 4. Kolmogorov and Cramer-von Mises distances under the alternative }\\
\multicolumn{8}{l}{\bf hypothesis when $n=m=100$ (4-cum only)} \\ \hline\hline
       &\multicolumn{3}{c}{K-dist}& &\multicolumn{3}{c}{CV-dist}\\ \cline{2-4} \cline{6-8}
Data   &Matching&Length &Counting & &Matching&Length &Counting\\ \hline\hline
Data I & 0.0076 &0.0132 &0.0176   & &0.0028  &0.0055 &0.0069\\
Data II& 0.0133 &0.0312 &0.0401   & &0.0045  &0.0101 &0.0065\\ \hline\hline
\end{tabular}\end{center}
\normalsize

\small\begin{center}\begin{tabular}{llclllllll}
\multicolumn{10}{l}{\bf TABLE 5. Comparison of probabilities in the left tail (4-cum only)}\\ \hline\hline
       &         &Sample& \multicolumn{7}{c}{Power (\%)} \\ \cline{4-10}
 Data  &  Measure&Size  & 50   & 60   & 70   & 80   & 90   &  95   & 99     \\ \hline\hline
       &         &    20& 48.59& 56.45& 65.41& 80.95& 92.40&  98.01& 100.00 \\
       & Matching&   100& 50.10& 59.63& 69.71& 79.27& 89.64&  95.62& 100.00 \\
       &         &  1000& 50.17& 60.10& 70.17& 79.89& 90.00&  95.00&  99.01 \\ \cline{2-10}
       &         &    20& 48.17& 57.12& 67.11& 78.42& 96.29&  99.63&  99.95 \\
Data II& Length  &   100& 50.00& 59.73& 69.26& 78.91& 89.48&  96.80&  99.91 \\
       &         &  1000& 50.13& 60.22& 70.13& 80.01& 89.97&  95.01&  99.06 \\ \cline{2-10}
       &         &    20& 48.41& 58.54& 67.59& 79.45& 96.01& 100.00& 100.00 \\
       & Counting&   100& 49.92& 59.79& 69.54& 79.19& 90.00&  97.12& 100.00 \\
       &         &  1000& 49.92& 59.92& 69.92& 79.95& 90.05&  94.99&  99.01 \\ \hline\hline
\end{tabular}\end{center}
\normalsize

\small\begin{center}\begin{tabular}{ccccccccc}
\multicolumn{9}{l}{\bf TABLE 6: Comparison of distances for (4-cum) chi-square and normal approximations}\\\hline\hline
        &        &      &\multicolumn{6}{c}{sample size $n$ ($m=3n$)}   \\ \cline{4-9}
Measure &Distance&Method&$  20$ &    50 &   100 &   500 &  1000 &  5000 \\ \hline\hline
        &K-dist  &Chi-sq&0.0288 &0.0187 &0.0116 &0.0047 &0.0077 &0.0068 \\
Matching&        &Normal&0.2030 &0.1325 &0.0915 &0.0408 &0.0324 &0.0144 \\ \cline{2-9}
        &CM-dist &Chi-sq&0.0154 &0.0096 &0.0059 &0.0022 &0.0028 &0.0025 \\
        &        &Normal&0.1494 &0.1021 &0.0694 &0.0314 &0.0237 &0.0085 \\ \hline\hline
        &K-dist  &Chi-sq&0.0269 &0.0163 &0.0054 &0.0072 &0.0074 &0.0093 \\
Length  &        &Normal&0.1779 &0.1160 &0.0805 &0.0365 &0.0267 &0.0099 \\ \cline{2-9}
        &CM-dist &Chi-sq&0.0127 &0.0079 &0.0020 &0.0027 &0.0035 &0.0035 \\
        &        &Normal&0.1191 &0.0805 &0.0541 &0.0248 &0.0147 &0.0054 \\ \hline\hline
        &K-dist  &Chi-sq&0.0246 &0.0174 &0.0090 &0.0078 &0.0087 &0.0070 \\
Counting&        &Normal&0.1721 &0.1112 &0.0757 &0.0333 &0.0233 &0.0127 \\ \cline{2-9}
        &CM-dist &Chi-sq&0.0122 &0.0085 &0.0036 &0.0029 &0.0040 &0.0033 \\
        &        &Normal&0.1089 &0.0694 &0.0456 &0.0208 &0.0161 &0.0084 \\ \hline\hline
\end{tabular}\end{center}\normalsize

\begin{center}\begin{tabular}{lrrrrrrrrr}
\multicolumn{10}{l}{\bf TABLE 7: Sample sizes required given significance level and power}\\ \hline\hline
            & \multicolumn{8}{c}{Power (\%)}\\ \cline{3-10}
            & \multicolumn{3}{c}{70}& \multicolumn{3}{c}{80}& \multicolumn{3}{c}{90}\\  \cline{3-4} \cline{6-7} \cline{9-10}
Significance (\%)& & CHB & JPT & & CHB & JPT & & CHB & JPT \\ \hline\hline
1        & & 181 & 186 & & 203 & 208 & & 234 & 240 \\
0.1       & & 275 & 282 & & 302 & 309 & & 339 & 348 \\
0.01      & & 366 & 375 & & 395 & 405 & & 438 & 449 \\
0.001     & & 435 & 446 & & 467 & 479 & & 513 & 526 \\ \hline\hline
\end{tabular}\end{center}

\end{document}